\documentclass{ws-mpla}
\usepackage[super]{cite}
\usepackage{hyperref}
\usepackage{graphicx}
\begin{document}

\markboth{A. Stepanian}
{On the invalidity of ``negative mass" description of the dark sector}

\catchline{}{}{}{}{}

\title{On the invalidity of ``negative mass" description of the dark sector\\}

\author{\footnotesize A. Stepanian}

\address{Center for Cosmology and Astrophysics, Alikhanian National Laboratory and Yerevan State University\\
 0036 Yerevan, Armenia,
\\
arman.stepanian@mail.yerphi.am}

\maketitle

\pub{Received (Day Month Year)}{Revised (Day Month Year)}

\begin{abstract}
It is shown that the concept of ``negative mass" introduced by Farnes (2018) to describe the dark sector within a unifying theory with the negative cosmological constant contradicts both the essence of the General Relativity and the available observational data. A viable model with modified weak-field General Relativity is mentioned. 
\end{abstract}

\ccode{PACS Nos.: 98.80.}

The paper \cite{UDS} aimed to describe the nature of  dark sector in a unified model, where the author's essential new notion is the ``negative mass''. That approach is already criticized in \cite{Soc}. Since, the ``negative mass" concept was publicized in media as a
new theory to explain the ``missing" part of cosmos (e.g. https://phys.org/news/2018-12-universe-theory-percent-cosmos.html), we will outline the shortcomings of that approach. 

First, the author claims that by considering the negative cosmological constant $\Lambda$ in General Relativity (GR) equations as the ``negative mass'' from one side, and in the modified Newtonian equation from the other side, it is possible to solve both the dark matter and dark energy problems, simultaneously. To do that, the author writes the equations of GR in the form
\begin{equation}\label{NMGR}
 G_{\mu\nu}=\frac{8 \pi G}{c^4} (T^{+}_{\mu\nu}+T^{-}_{\mu\nu}+C_{\mu\nu}),
\end{equation}
where according to the author ``\textit{... the conventional $\Lambda g_{\mu\nu}$ term is now represented by a combination of $T^{-}_{\mu\nu}$ (an exotic matter term) and $C_{\mu\nu}$ (a modified gravity term)}''. So, in order to explain the accelerated expansion of the universe, the author replaces the notion of positive cosmological constant $\Lambda$ with a new term in the right hand side of the field equations above. This is severe misunderstanding of GR since the author's statement entirely rejects the very essence of Einstein equations as explained in any textbook!\footnote{See e.g. the strong critics by t'Hooft  \cite{t'Hooft} of such attempts to interpret the Einstein equations.} That very essence of Einstein equations has been triumphantly confirmed by now experimentally, it is enough to mention the discovery of gravitational waves \cite{Ab} and the frame dragging measurements \cite{Ciu}, testing the strong and weak-field GR, respectively.     

Second, in the Newtonian regime, the author obtains the following relation for the gravitational force exerted on a particle of mass $m$
\begin{equation}\label{NL}
\frac{m v^2}{r}=-\frac{GM m}{r^2}+\frac{\Lambda c^2 r m}{3}.
\end{equation}
To ensure the ``negative mass'' as the ``missing matter'' in galaxies i.e. to ensure flat rotation curves, the author claims that the negative cosmological constant has to be 
$$\Lambda \simeq -0.3\times 10^{-52} m^{-2}.$$ 
This contradicts a variety of observational data including the SN surveys, CMB, lensing etc when interpreted via cosmological constant and GR. Namely, the observations indicate the positive $\Lambda$ and fix its value  $\Lambda \simeq +1.1 \times 10^{-52} m^{-2}$, as of the Planck data \cite{Pl}. Furthermore, with the negative value of $\Lambda$ there will be in contradiction with the gravitational potential
\begin{equation}\label{PotL} 
\Phi=-\frac{GM}{r}-\frac{\Lambda c^2 r^2}{6},
\end{equation}
which is equally important on introducing the notion of ``missing matter''. In fact, it should be noticed that there are two main analyses for the existence of dark matter i.e. of the ``virial theorem'' and the ``flat rotation curves''. Historically, the concept of ``missing matter'' was introduced by Zwicky \cite{ZW} using the virial theorem. Decades later this concept was used to describe the ``flatness" of galactic rotation curves \cite{VR}. In contrast to virial analysis where the gravitational potential is used, to study the rotation curves of galaxies one uses the gravitational force and there is an important difference between those analyses. Namely, if one considers the additional $\Lambda$-term in Eq.(\ref{PotL}), then the  $\Lambda$-term will correspond to ``mass'' term in Eq.(\ref{PotL}) with density

\begin{equation}\label{density}
\rho= \frac{\Lambda c^2}{8 \pi G}.
\end{equation}

While for Eq.(\ref{NL}), in view of $F_g =m E_g= -m\nabla \Phi$, the sign of  $\Lambda$-term changes and it should be interpreted as ``negative mass'' of
\begin{equation}\label{density2}
\rho= -\frac{\Lambda c^2}{4 \pi G}.
\end{equation}

In \cite{UDS} to solve the dark matter problem the author has focused only on the ``rotation curve'' analysis based on Eq.(\ref{NL}) and has claimed that considering the ``negative mass'' defined via the negative $\Lambda$ it is possible to have flat rotation curves for galaxies. In fact, although it will act to flatten the rotation curves, the negative $\Lambda$  term cannot explain the other main analysis i.e. of the``virial theorem''. Indeed, in such case the $\Lambda$-term in the gravitational potential will become positive
\begin{equation}\label{PotL1} 
\Phi=-\frac{GM}{r}+\frac{|\Lambda| c^2 r^2}{6},
\end{equation}
where  $|\Lambda|$ is the absolute value of the cosmological constant. Consequently, for the virial theorem the ``negative mass" not only will not help but even will decrease the amount of necessary mass to have a virialized configuration.

Thus, what the author calls a ``unifying theory'' for dark energy and dark matter via negative value of $\Lambda$ actually is just a matter of reparametrization of $\Lambda$, c and G based on Eq. (\ref{NL}) (cf. \cite{KEK}),  which has its own theoretical and observational constraints, as it is clear comparing Eq.(\ref{density}) with Eq.(\ref{density2}) and recalling that both the sign and the value of $\Lambda$ are confirmed by various observations.

Third, the author claims that Anti de Sitter (AdS) universe undergoes a cycle of expansion and contraction with a timescale of $$\sqrt{\frac{-3\pi^2}{\Lambda c^2}},$$ where the author states that  ``\textit{...a universe with a negative cosmological constant would eventually recollapse due to this extra attractive force''}. However, negative cosmological constant is neither necessary nor sufficient condition to have a Big Bang/Big Crunch scenario and AdS spacetime is totally unrelated to a  collapsing universe and the negative $\Lambda$ cannot ensure any so-called ``cyclic cosmology'', see e.g. \cite{GV}.

Finally, let us mention that there is still a way to incorporate the cosmological constant within Newtonian gravity without rejecting basic principles and observations, namely via Newton theorem on the equivalency of attractions of a sphere and a point mass. The latter leads to the modification of the weak-field limit of GR as follows \cite{GS,GS1}
\begin{equation}\label{WF}
g_{00} = 1 - \frac{2 G m}{r c^2} - \frac{\Lambda r^2}{3};\,\,\, g_{rr} = (1 - \frac{2 G m}{r c^2} - \frac {\Lambda r^2}{3})^{-1}.
\end{equation}
Then, the cosmological constant not only enters naturally in the gravity equations, both Newtonian and GR, but also enables one to describe via the observed value of the cosmological constant the common nature of the dark sector, without any need to change the sign of $\Lambda$.

In sum, the ``negative mass" and negative cosmological constant in \cite{UDS} contradicts basic physics and a bunch of observational data. 

\section*{Acknowledgement}
I am thankful to the referees for valuable comments. This work is partially supported by the ICTP through AF-04.


\begin{thebibliography}{0}

\bibitem{UDS} J.S. Farnes, {\it A}\&{\it A}, {\bf A92}, 620,  (2018). 

\bibitem{Soc} H. Socas-Navarro, {\it A} \&{\it A}, {\bf A5}, 626, (2019).

\bibitem{t'Hooft} G.'tHooft, \url{http://www.staff.science.uu.nl/~hooft101/gravitating\_misconceptions.html}

\bibitem{Ab} B.P. Abbott , et al, {\it Phys. Rev. Lett.}, {\bf 061102},  116, (2016).

\bibitem{Ciu} I. Ciufolini , et al, {\it Eur. Phys. J.  C}, {\bf 76},  120,  (2016).

\bibitem{Pl} P.A.R. Ade, et al, {\it A\& A}, {\bf A13} 594,  (2016).  

\bibitem{ZW} F. Zwicky,  {\it Helv. Phys. Acta.}, {\bf 6}, (1933). 

\bibitem{VR} V. Rubin,  K. Ford,  {\it ApJ}, {\bf 159}, (1970).

\bibitem{KEK} S. Kopeikin,  M. Efroimsky, G. Kaplan, {\it Relativistic celestial mechanics of the solar system}, (Wiley,  2001).

\bibitem{GV} M. Gasperini, G. Veneziano, {\it Phys. Rep.}, {\bf 1}, 373, (2003). 

\bibitem{GS} V.G. Gurzadyan,  A. Stepanian, {\it Eur Phys. J. C}, {\bf 78},  632, (2018).

\bibitem{GS1} V.G. Gurzadyan,  A. Stepanian,  {\it Eur Phys. J. C},  {\bf 79}, 169, (2019).

\end{thebibliography}
\end{document}